\documentclass[
aps,superscriptaddress,
twocolumn,prd,
showpacs,preprintnumbers,nofootinbib,
amsmath,amssymb,
aps,
floatfix,
superscriptaddress]{revtex4-1}
\usepackage[printonlyused]{acronym}
\usepackage{comment}
\usepackage{graphicx, epsfig, amssymb}
\usepackage{amsmath, amsfonts}
\usepackage{gensymb}
\usepackage{bm}
\usepackage[linktocpage]{hyperref}

\newcommand{\mnras}{Monthly Notices of the Royal Astronomical Society}

\newcommand{\apjl}{Astrophysical Journal Letters}

\makeatletter

\usepackage{url}

\begin{document}

\title{GW190521 as a black-hole merger coincident with the ZTF19abanrhr flare}

\author{Juan Calder\'{o}n Bustillo}
 	\email{juan.calderon.bustillo@gmail.com}
	\affiliation{Instituto Galego de F\'{i}sica de Altas Enerx\'{i}as, Universidade de
Santiago de Compostela, 15782 Santiago de Compostela, Galicia, Spain}
	\affiliation{Department of Physics, The Chinese University of Hong Kong, Shatin, N.T., Hong Kong}

\author{Samson H.W. Leong}
	\affiliation{Department of Physics, The Chinese University of Hong Kong, Shatin, N.T., Hong Kong}

\author{Koustav Chandra}
	\affiliation{Department of Physics, Indian Institute of Technology Bombay, Powai, Mumbai, Maharashtra 400076, India}
    
\author{Barry McKernan}
	\affiliation{Department of Science, CUNY-BMCC, 199 Chambers St., New York, NY 10007, USA}
    \affiliation{Department of Astrophysics, American Museum of Natural History, Central Park West, New York, NY 10028, USA}
    \affiliation{Physics Program, The Graduate Center, CUNY, New York, NY 10016, USA}
    \affiliation{Center for Computational Astrophysics, Flatiron Institute, New York, NY 10010, USA}
\author{K. E. S. Ford}
	\affiliation{Department of Science, CUNY-BMCC, 199 Chambers St., New York, NY 10007, USA}
    \affiliation{Department of Astrophysics, American Museum of Natural History, Central Park West, New York, NY 10028, USA}
    \affiliation{Physics Program, The Graduate Center, CUNY, New York, NY 10016, USA}
    \affiliation{Center for Computational Astrophysics, Flatiron Institute, New York, NY 10010, USA}
\begin{abstract}

We present an analysis that reconciles the \acl{GW} signal GW190521 observed by the Advanced LIGO and Advanced Virgo detectors with the electromagnetic flare ZTF19abanrhr observed by the Zwicky Transient Facility. We analyze GW190521 under a mass-ratio prior uniform in $Q \in [1,4]$ using the state-of-the-art waveform model for black-hole mergers \texttt{NRSur7dq4}. We find a $90\%$ credible region for the black-hole masses extending far outside what originally reported by \cite{GW190521D}, where our maximum likelihood masses reside. We find a $15\%$ probability that both \aclp{BH} avoid the pair-instability supernova gap. We infer a three-dimensional sky-location highly consistent with ZTF19abanrhr, obtaining an odds-ratio ${\cal{O}}_{C/R}=72:1$ that strongly favors the hypothesis of a true coincidence over a random one. Combining this event with the neutron-star merger GW170817, we estimate a Hubble constant H$_0=72.1^{+10.6}_{-6.4}\mathrm{km\,s^{-1}\,Mpc^{-1}}$ at the $68\%$ credible level. 

\end{abstract}

\maketitle

\acrodef{BBH}[BBH]{binary black hole}
\acrodef{BH}[BH]{black hole}
\acrodef{LVK}[LVK]{LIGO, Virgo and KAGRA collaborations}
\acrodef{GW}[GW]{gravitational wave}
\acrodef{CI}[CI]{credible interval}
\acrodef{PSD}[PSD]{power spectral density}


\section{Introduction} The gravitational-wave (GW) detectors Advanced LIGO \cite{aLIGO}, and Advanced Virgo \cite{aVirgo} have made the observation of compact binary mergers almost routine. In only six years, these have reported $\sim \mathcal{O}(90)$ such observations \cite{GWTC3, GWTC2, GWTC1}. These have provided us with unprecedented knowledge on how BHs and neutron stars form and how they populate our Universe. Moreover, these observations have enabled the first tests of General Relativity in the strong-field regime, and qualitatively new studies of the Universe at a large scale \cite{Abbott:2016blz, GWTC1}. Unleashing such scientific potential from GW observations requires accurate inference of the source properties. This has been largely possible for most observations owing to both accurate, computationally efficient waveform models and to the fact that most detections displayed relatively long pre-merger, inspiral stages that provided us with information about the individual merging bodies.


The detection of GW190521 by the \acf{LVK} represented the first departure from such ``canonical'' events \cite{GW190521D, GW190521I}. Owing to the large mass of its source, GW190521 barely displays any pre-merger dynamics, with the vast majority of the signal coming from the final distorted, merged object as it relaxes to its final BH form. In such a situation, there is little information about the parents of the final object, causing the inference of the source parameters to depend strongly on prior assumptions \cite{IAS_GW190521, GW190521_Nitz, Isobel_ecc, Gayahtri_ecc, GW190521_Gamba,Proca}. This has led to a large variety of interpretations of this event.

First, the LVK reported a quasi-circular \ac{BBH} merger with signatures of orbital precession involving at least one BH populating the pair-instability supernova (PISN) gap \cite{Fryer:2000my, Heger:2002by}. Second, using a population-informed mass prior, \citet{straddling} hinted that GW190521 could involve one black-hole above the PISN gap and one below, known as a ``straddling binary''. Next, \cite{HOC} showed that for such short signals, orbital precession could be confused with high eccentricity. Consistently, \citet{Isobel_ecc} and \citet{Gayahtri_ecc} showed that GW190521 is also consistent with an eccentric merger and \citet{GW190521_Gamba} even pointed to the possibility of a dynamical capture \cite{Nagar2021}. More important for this work, \citet{GW190521_Nitz} showed that using a mass-ratio prior uniform in $Q =m_1/m_2 \geq 1$ could lead to an interpretation as an intermediate high-mass ratio BBH; and similar results were found by \citet{Hector_GW190521}. Finally, \cite{Proca} showed that the event is even consistent with the merger of horizonless exotic compact objects known as Proca stars \cite{brito2016proca, sanchis2019head}.

GW190521 could also be the first multi-messenger observation of a \ac{BBH} event. \citet{flare}, reported the observation of an electromagnetic signal, ZTF19abanrhr by the Zwicky Transient Facility (ZTF) \cite{ZTF_1, ZTF_2} in a region of the sky consistent with that initially reported by the \ac{LVK} in an early warning~\cite{21g_GCN}, proposing it as a counterpart to GW190521. If true, this would have paramount implications in interpreting GWs from compact mergers, forecasts for future counterparts and measurements of the Hubble constant. However, due to larger inconsistencies with the sky-location finally estimated by the LVK for GW190521 \cite{GW190521D}, \citet{Greg} showed that a true association was unlikely. The same was concluded by \cite{GW190521_Nitz,Hector_GW190521} under their analyses with alternative mass priors. Importantly, however, while GW190521 is a merger-ringdown dominated signal with signatures of precession, \cite{Hector_GW190521} (and in part \cite{GW190521_Nitz}) compared GW190521 to phenomenological waveform models \cite{XPHM_Pratten,XPHM_Cecilio,XPHM_Cecilio_2,TPHM_THM,TPHM_22,TPHM_HM} that are calibrated to numerical simulations of BBHs with no orbital precession, modelling this through analytical approximations that break during the merger-ringdown regime \cite{Schmidt:2012rh,Hannam:2013oca} (See Appendix~\ref{sec:maxL}).\\

Here we analyse GW190521 under several mass priors using the state-of-the-art waveform model \texttt{NRSur7dq4} \cite{NRSur7dq4}. This model, labelled as ``preferred'' by the LVK in their original analysis of GW190521 \cite{GW190521D,GW190521I}, is directly calibrated to 1528 numerical simulations of precessing BBHs \cite{SXS} with mass ratios $Q=m_1/m_2 \leq 4$, therefore including all the physics present in these systems. Imposing a mass-ratio prior uniform in $Q=m_1/m_2\in [1,4]$, we find that GW190521 is consistent with a \ac{BBH} with masses $m_1=106^{+36}_{-28}\,M_\odot$ and $m_2=61^{+21}_{-19}\,M_\odot$ at a distance of $2.8^{+3.4}_{-1.7}$\,Gpc.
Considering a PISN gap spanning the mass range $[65,130]\,M_\odot$, we find a $15\%$ probability that GW190521 is a ``straddling binary''. Moreover, we recover a sky-location highly consistent with ZTF19abanrhr. Reproducing the analysis in \citet{Greg}, we obtain an odds of ${\cal{O}}_{C/R}=72$ against a random spatial coincidence, indicative of a \textit{strong evidence for a true association}~\cite{KassRaftery}. We note that while  \cite{GW190521_Nitz} did also use the model \texttt{NRSur7dq4} they extended it past its calibration region, up to $Q\leq 6$. Doing so we find a lower but still strong odds-ratio of ${\cal{O}}_{C/R}=47$ (see also Appendix A).  

\section{Analysis setup} 

\paragraph{\textbf{Data and waveform model}} We perform full Bayesian parameter inference on 4 seconds of publicly available data \cite{GWOSC} from the Advanced LIGO \cite{aLIGO} and Advanced Virgo \cite{aVirgo} detectors around the time of GW190521 sampled at 1024\,Hz and with a subtraction of the 60\,Hz US-power line \cite{Derek_subtraction, Viets:2021aaa}. We employ the same power-spectral-density estimate as in \cite{GW190521D}. We compare GW190521 to theoretical waveform templates for \ac{BBH} predicted by General Relativity generated with the waveform model \texttt{NRSur7dq4} \cite{NRSur7dq4}. This model is calibrated to numerical simulations of precessing quasi-circular \ac{BBH}s with mass ratios $Q \leq 4$ and dimensionless spin-magnitudes $a_i \leq 0.8$ but can be extrapolated up to $Q\leq 6$ and $a_i \leq 0.99$.\\ 
\paragraph{\textbf{Priors}} Following standard practice, in their original analysis of GW190521, the LVK  placed uniform priors on the red-shifted component masses \cite{GW190521D, GW190521I}. We will refer to this as the LVK prior. This, however, translates into a strong prior for rather equal-mass systems that severely punishes high mass ratio regions where the best-fitting parameters may reside \cite{GW190521_Nitz}. Here, we test two different mass-ratio priors: one uniform in $q=m_2/m_1 \in [0.25,1]$, that also favours equal-mass ratios, and another one uniform in $Q=m_1/m_2 \in [1,4]$ that removes such preference. Throughout the text, we will respectively refer to these as $Q$ and $q$-runs. As a consistency test, since the \texttt{NRSur7dq4} model can be extrapolated to $Q\leq 6$, we perform additional runs extending our priors to respectively $q=1/6$ and $Q=6$. We place a uniform prior on the total red-shifted mass and a prior on luminosity distance uniform in co-moving volume assuming Planck 2015 cosmology. In contrast, the \ac{LVK} employed a prior $p(d_L) \propto d_L^2$, uniform in Euclidean Volume. Finally, we place standard priors in all remaining parameters. Consistently with the LVK analysis \cite{GW190521D}, we compute spin-values at a fiducial reference frequency of $f_\text{ref}=11$\,Hz and use a minimum frequency cutoff of $f_\text{min}=11$\,Hz. 
\\

\paragraph{\textbf{Sampler Settings}} We employ the parameter estimation software \texttt{Parallel Bilby}~\cite{pbilby,Ashton:2018jfp} equipped with the nested sampler \texttt{Dynesty} \cite{Dynesty}, using 4096 live points. In contrast, we note that the LVK analysis was performed with the LALInference software \cite{Veitch:2015ela} using its nested sampling scheme \texttt{LALInferenceNest} \cite{LalinferenceNest} with 2048 live points.\\

\paragraph{\textbf{LVK cross-check}} As a consistency check, we also reproduce the LVK analysis by placing the same mass and distance priors.\\

\section{Results}
Table~\ref{tab:PE} summarises our results, contrasting them with those of the LVK. We report median values and symmetric 90$\%$ credible intervals (CIs) except for  Hubble constant H$_0$, for which following standard practice, we report $68\%$ CIs. The first two columns report results corresponding to our $Q$-analyses. Throughout the text, we will use the first column $(Q\leq 4)$ as our ``preferred'' results, with those in the second column $(Q\leq6)$ serving as a robustness check\footnote{We note that in this analysis, we find a probability $p(Q\geq 4) < 0.03$.}. Results in the central columns correspond to our $q$-analyses, and the last two columns show the result of our reproduction of the LVK analysis and the original LVK result, respectively. 

First, we note that our reproduction of the LVK analysis matches well with their result\footnote{We attribute the small difference in maximum SNR to the different number of sampling live points (4096 vs 2048).}. Second, as expected, $q$-analyses also report results consistent with the LVK ones. Finally, results from both $Q$-analyses are self-consistent but deviate from the other four.\\

\begin{figure}[ht]
\centering
\includegraphics[width=0.492\textwidth]{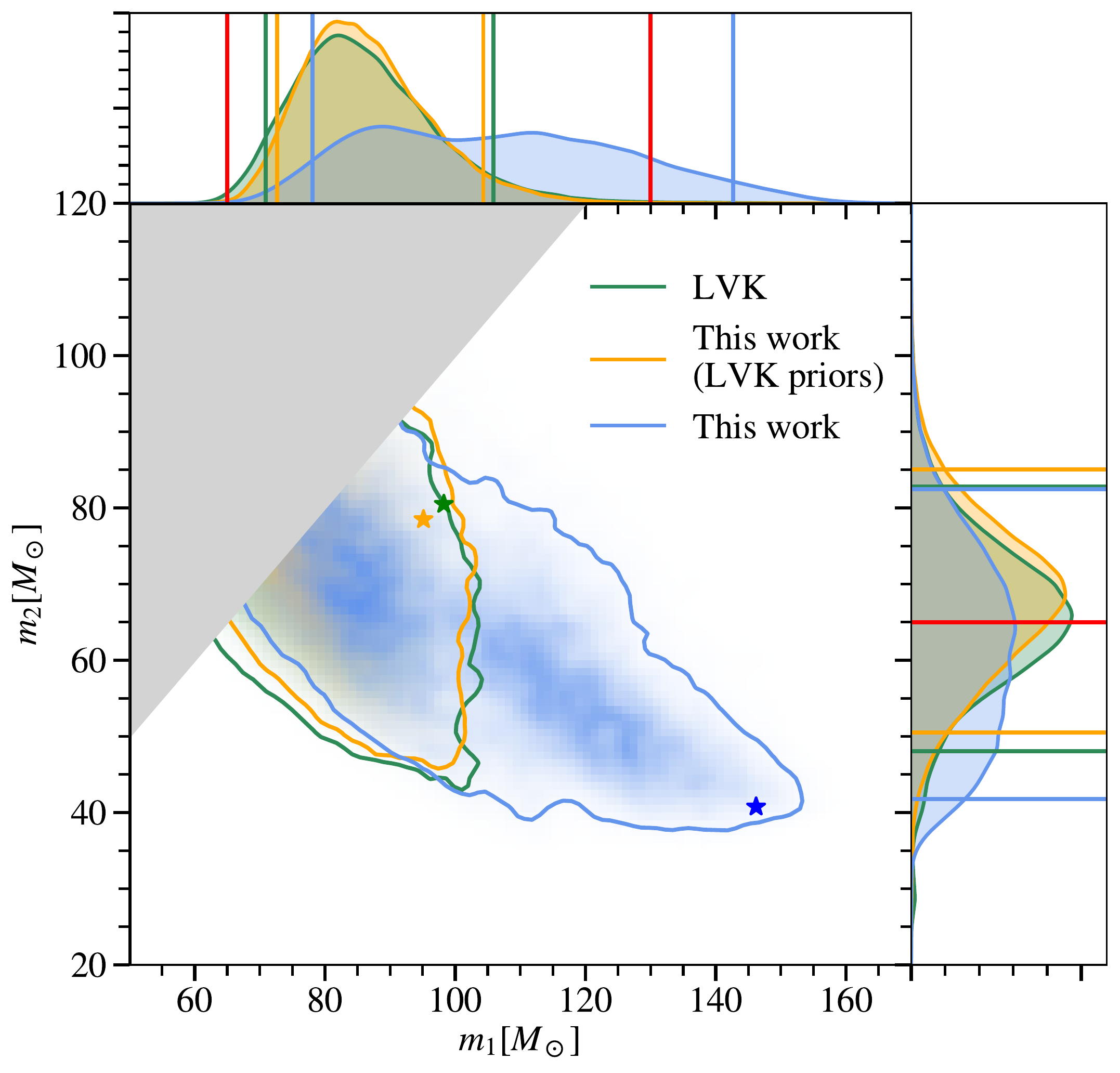}
\caption{\textbf{Posterior distributions of the component masses of GW190521}. The blue, green, and orange contours denote the $90\%$ credible regions respectively obtained by us using a mass-ratio prior uniform in $Q\in[1,4]$, by the LVK analysis and by us when using the LVK prior settings. The respective lines in the side plots denote symmetric $90\%$ credible intervals around the median, while the red ones denote the limits of the PISN gap. The stars denote the maximum likelihood masses.}
\label{fig:masses}
\end{figure}
\paragraph{\textbf{Model selection}} 
Fig.~\ref{fig:masses} shows the posterior distributions for the component masses of GW190521 according to three different analyses. In blue is our preferred analysis, while green and orange respectively denote the original LVK analysis and our reproduction of it. The contours in the main panel represent the corresponding two-dimensional $90\%$ credible regions, with the three stars denoting the masses of the respective maximum likelihood ($\max\mathcal{L}$), best-fitting templates. We show these in Appendix \ref{sec:maxL} overlaid onto the detector data. The colour darkness denotes the probability density for our preferred analysis. On the sides, we show the corresponding distributions for the two masses. We delimit the $90\%$ credible intervals with vertical bars. Finally, red lines denote the ends of the PISN gap.

First, as expected from Table~\ref{tab:PE}, we note that green and orange contours overlap well and that the corresponding $\max\mathcal{L}$ masses lay close to each other. This, together with Table~\ref{tab:PE}, indicates that we correctly reproduce the \ac{LVK} results. Second, while our blue contour encompasses the LVK one, it also contains a secondary region of high probability density that extends to larger mass-ratio regions, with our $\max\mathcal{L}$ masses residing deep in such region. Third, while the Bayes Factors in the bottom row of Table \ref{tab:PE} show \textit{no strong preference for any of these analyses}, we note that our best-fitting waveform yields a much larger $\max\mathcal{L}$ and signal-to-noise ratio (SNR) than both the LVK analysis and our reproduction of it. In particular, the latter yields a $\ln(\max\mathcal{L})=114.1$ that is 5.1 units smaller than the $\ln(\max\mathcal{L})=119.2$ we obtain \footnote{This is, our best-fit template is $e^{5.1}\simeq 164$ times more likely given the detectors data.}. Similarly, $q$-analyses also report lower $\max\mathcal{L}$. This showcases that, while not discarded by their Bayes' Factors, these analyses can miss even the highest-likelihood regions of the parameter space due to their prior settings, potentially missing important information about the source. We understand that this, together with the lack of observational constraints in this region of the parameter space, justifies the investigation of alternative priors that can explore such regions and the following detailed analysis of the corresponding results.\\ 

\begin{figure}[ht]
\centering
\includegraphics[width=0.492\textwidth]{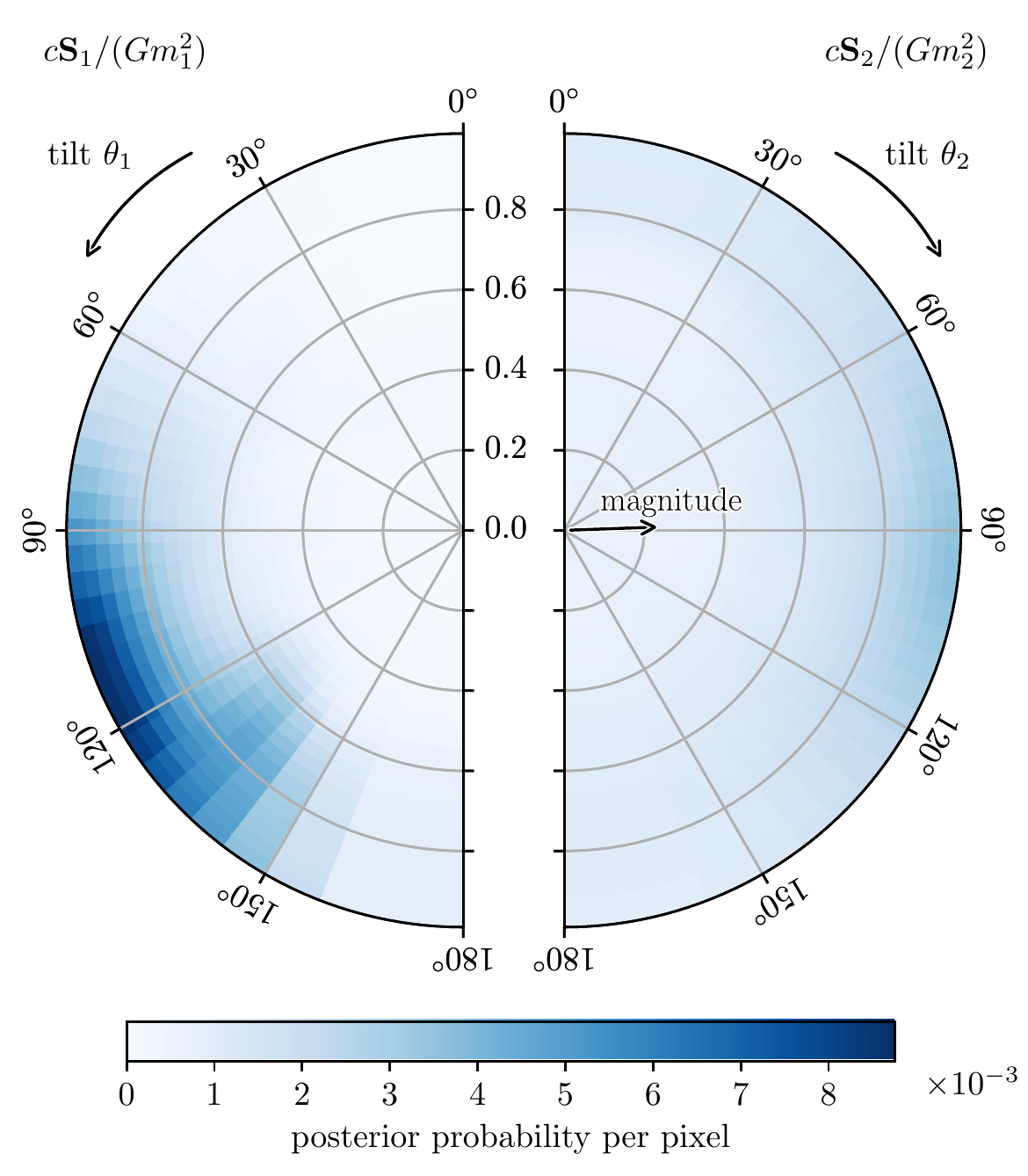}
\caption{\textbf{Posterior distributions for the magnitude and tilt of the GW190521 spins}. The radial coordinate represents the dimensionless spin magnitudes $a_{1,2}$. The angle denotes the spin tilts $\theta_{1,2}$, defined as the angle between the spin and the orbital angular momentum of the binary at a reference frequency $f_\text{ref}=11$\,Hz. A zero tilt denotes alignment with the orbital angular momentum. Non-zero tilts induce orbital precession. All pixels have equal prior probability. When replicating the LVK setup, we obtain results consistent with those in \cite{GW190521D}.}
\label{fig:spindisk}
\end{figure}

\paragraph{\textbf{Source parameters}} We infer individual masses $m_1=106^{+36}_{-28}\,M_\odot$ and $m_2=61^{+21}_{-19}\,M_\odot$. This yields a mass ratio $Q=1.71^{+1.50}_{-0.65}$ and total mass $M_{\text{tot}}=168^{+29}_{-27}\,M_\odot$, both larger than those reported in \cite{GW190521D}. If we assume that the PISN gap spans the range $[65,130]\,M_\odot$, the primary \ac{BH} has a $15\%$ probability of being above the gap while the secondary has a $65\%$ probability of residing below it. The probability that both \acp{BH} are out of the gap is $13\%$, in high contrast with the $0.001\%$ reported in \cite{GW190521D}. We also infer a larger orbit inclination $\iota=70^{+17}_{-45}$ deg than reported in \cite{GW190521D}. Altogether this translates into a much weaker source (See Appendix~\ref{sec:loudness}) that yields a significantly closer distance of $2.8^{+3.4}_{-1.7}\,$Gpc and a smaller redshift $z=0.49^{+0.44}_{-0.26}$ that peaks at the value of ZTF19abanrhr (see later). 

Fig.~\ref{fig:spindisk} shows the posterior distribution for the magnitude and the tilt of spins w.r.t. orbital angular momentum. Because our mass-ratio posterior extends to larger values than the LVK, we find the data to be more informative about the primary spin but we can barely retrieve information about the secondary~\citep{Biscoveanu:2021nvg}. Similar to \cite{GW190521D}, we find that, while largely unconstrained, the spin magnitudes $a_{1,2}$ peak near the Kerr limit $a=1$. We constrain the tilt of the primary spin to $\theta_1 = 107^{+36}_{-53}$ deg. Tilted spins -- i.e. with non-zero components within the orbital plane -- induce orbital precession via spin-orbit coupling \cite{Apostolatos:1994mx,Kidder:1995zr}. The impact of precession in the waveform is captured by effective precessing spin $\chi_p$~\cite{Schmidt:2012rh,Hannam:2013oca}, for which we obtain a posterior distribution consistent with the LVK. Finally, we obtain a $71\%$ probability for the off-plane component to be anti-aligned with the orbital angular momentum. Combining this with the barely informative posteriors on the secondary spin yields a lower effective-spin parameter~\cite{Santamaria:2010yb,Ajith:2009bn} than reported by the LVK of $\chi_{\text{eff}} = -0.13^{+0.37}_{-0.33}$\footnote{The effective-spin is defined as the mass-weighted average of the projections of the black-hole spins $\vec{a_i}$ along the orbital angular momentum as $\chi_{\text{eff}}=\frac{a_1^z m_1+a_1^z m_2}{m_1+m_2}$. The effective-precessing spin is defined as $\chi_p = \text{max}(a_1^{\perp},a_2^{\perp}(A/q^2))$ with $A=2+1.5q$ and $a_i^{\perp}$ denoting the projection of $\vec{a_i}$ onto the orbital plane}.

We use the \texttt{SurfinBH} package \cite{SurfinBH} and the fit \texttt{NRSur7dq4Remnant}~\cite{NRSur7dq4} to estimate the properties of the remnant \ac{BH} from the BBH ones. We obtain a final spin $\chi_f=0.65^{+0.12}_{-0.23}$ and a final mass $M_f = 162^{+29}_{-28}\,M_\odot$ with zero support below $100\,M_\odot$ thus maintaining its original interpretation as an intermediate-mass black hole~\cite{GW190521D}.\\

\begin{figure}[t]
\centering
\includegraphics[width=0.492\textwidth]{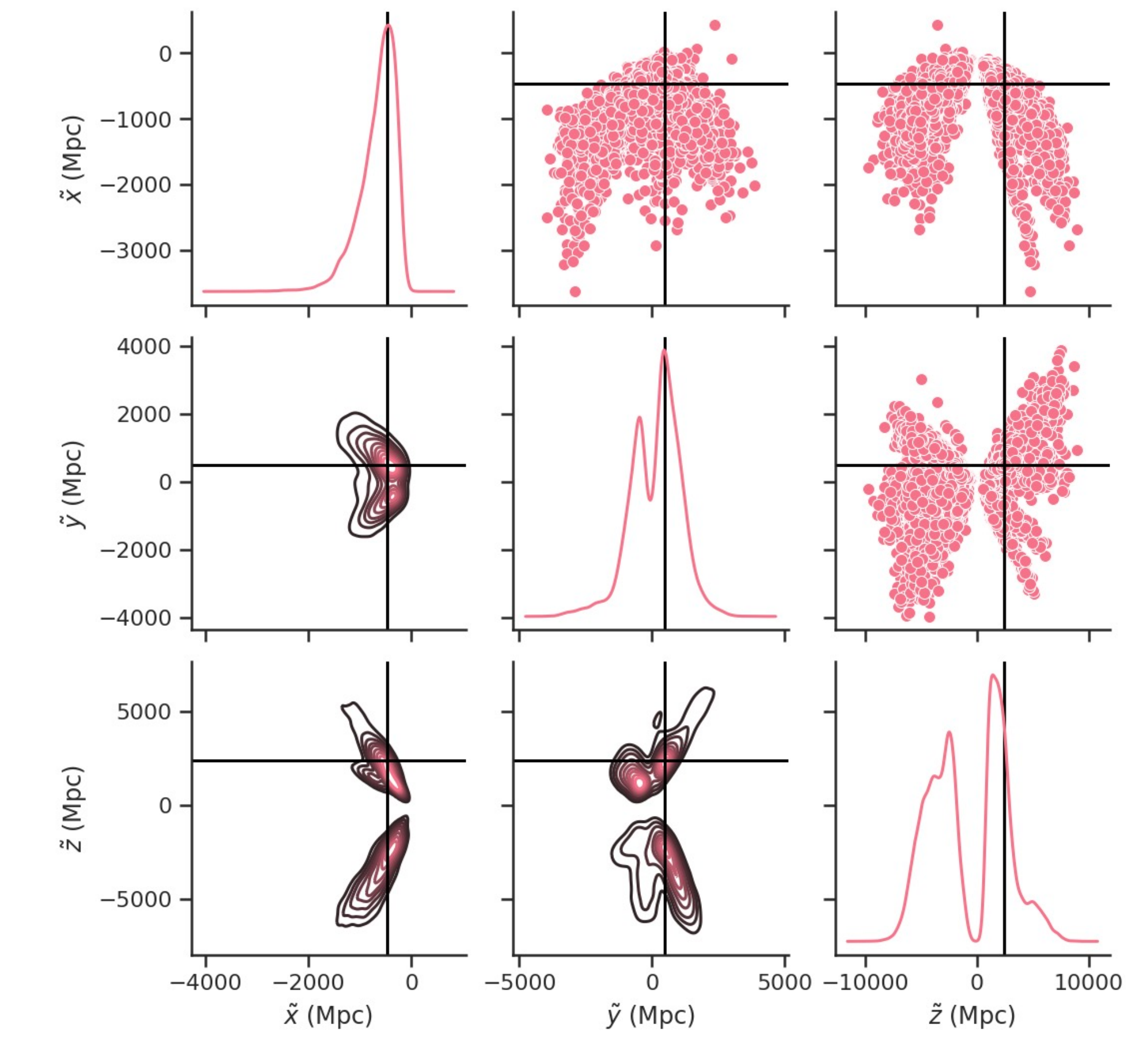}
\caption{\textbf{Two-dimensional and one-dimensional posterior distribution for the location of GW190521 in Cartesian coordinates}. The black lines denote  the location of ZTF19abanrhr. The coordinates are defined in the principal axes of the skymap.}
\label{fig:xyz}
\end{figure}

\begin{figure}[t]
\centering
\includegraphics[width=0.49\textwidth]{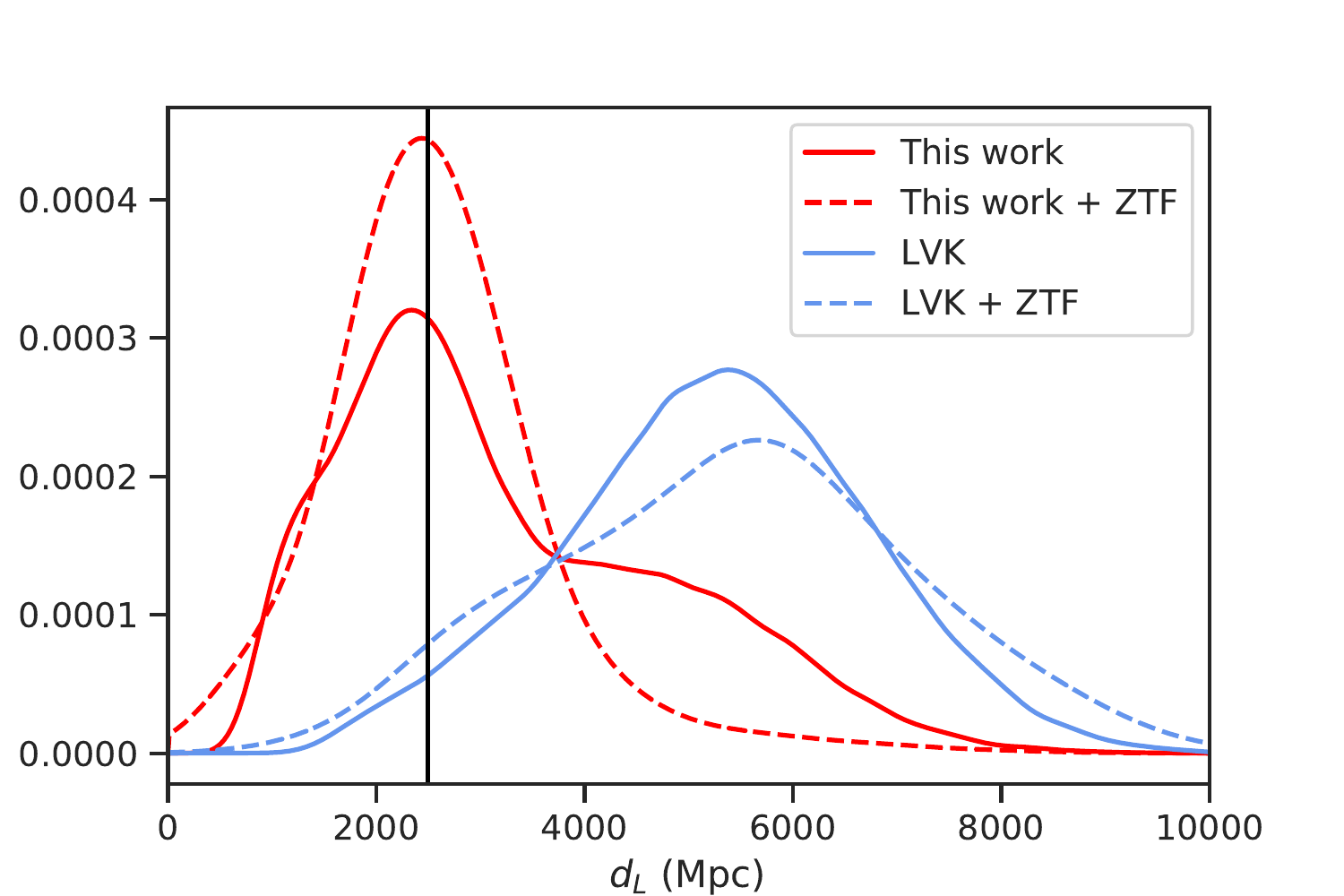}
\caption{\textbf{Posterior distributions for the luminosity distance of GW190521}. Solid lines show the full posterior distributions, while dashed ones show those conditioned to the sky-location of ZTF19abanrhr, whose distance is denoted by the black vertical line.}
\label{fig:distances}
\end{figure}

\begin{figure*}[t]
\begin{center}
\includegraphics[width=0.32\textwidth]{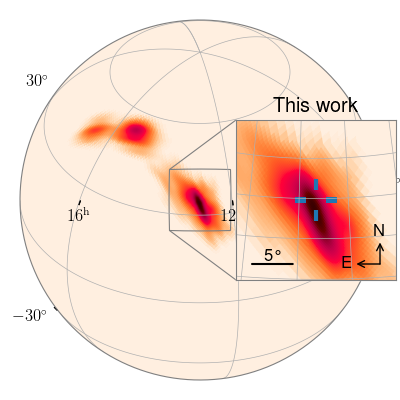}
\includegraphics[width=0.32\textwidth]{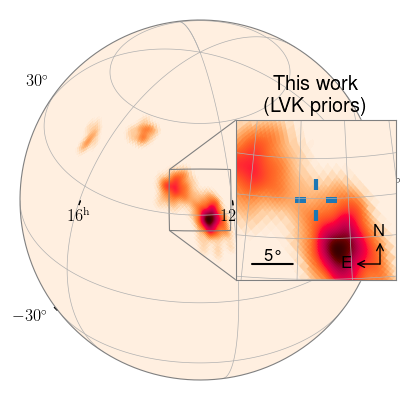}
\includegraphics[width=0.32\textwidth]{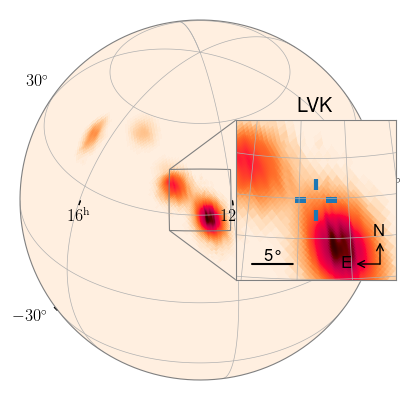}
\caption{\textbf{Sky-location of the flare ZTF19abanrhr \cite{flare} and GW190521.} Posterior probability densities for the sky-location of GW190521 according to the LVK analysis \cite{GW190521I} (right) and to ours (left). As a cross-check, the central panel shows the result of our reproduction of the LVK analysis. The blue cross represents the location of ZTF19abanrhr.}
\label{fig:skymaps}
\end{center}
\end{figure*}

\paragraph{\textbf{Association with ZTF19abanrhr}} Fig.~\ref{fig:xyz} shows the one and two-dimensional posterior distributions for the three-dimensional location of GW190521. This is expressed in terms of the coordinates $(\tilde{x},\tilde{y},\tilde{z})$ describing the principal axes of the sky-map. All distributions peak close to the location of ZTF19abanrhr, given by $(\text{RA},\,\text{DEC})=(192.42625^\circ,\,34.82472^\circ)$ and a redshift $z=0.438$ (black lines). To provide a more common visualization and to facilitate comparison with the results of the LVK, Fig.~\ref{fig:skymaps} shows the posterior probability distributions for the sky-location of GW190521 for our analysis (left), LVK analysis (centre) and our reproduction of the latter (right). In addition, Fig.~\ref{fig:distances} shows the corresponding posterior distributions for the luminosity distance together with those conditional to the sky-location of ZTF19abanrhr (dashed). In all cases, our distributions (red) peak near the distance value of ZTF19abanrhr, which falls in the tail region of the LVK ones (blue).

Next, we assess the probability of a true coincidence. Using the formalism in \cite{Ashton_coninc_stat,Greg} we compute the odds of a common-source hypothesis, $C$, against a random coincidence, $R$, as ${\cal{O}}_{C/R}=\pi_{C/R} \times \mathcal{I}_\theta$. Here, $\mathcal{I}_\theta$ (or three-dimensional overlap integral) denotes the ratio between the posterior probability density (derive from GW190521 data $d_\text{GW190521}$) of the sky-location and distance, and the common-source hypothesis prior; both evaluated at the parameter values $\theta_\text{ZTF}$ of ZTF19abanrhr. This is
\begin{equation}
\mathcal{I}_\theta=\frac{p(\theta_{\rm ZTF} \mid d_{\text{GW190521}},\,C)}{\pi (\theta_{\rm ZTF} \mid C)},
\label{eq:greg}
\end{equation}
where $\theta=\{\text{RA},\text{DEC},d_L\}$. The factor $\pi_{C/R}$ denotes the prior odds against a true coincidence which,  as in~\cite{Greg}, we set this to $\pi_{C/R} = 1/13$, which is determined by the number of flares similar to ZTF19abanrhr in the ZTF alert stream. Finally, as in ~\cite{Ashton:2018jfp}, we re-weight~\cite{Reweghting} our posterior samples to consider a distance prior uniform in co-moving volume and source-frame time \cite{Isobel-catalogue} $p(z)\propto dV(z)/[d z (1+z)]$.\\ 

For our reproduction of the LVK analysis, we obtain a weak evidence ${\cal{O}}^{\text{LVK, ours}}_{C/R} = 2.9$, qualitatively equivalent to the value ${\cal{O}}^{\text{LVK}}_{C/R} = 2.4$ obtained by \cite{Greg}.
In contrast, due to the much better sky-location and distance agreement from our $Q$-analysis, we obtain a strong odds of ${\cal{O}}_{C/R} = 72$. To check the robustness of this result, we repeated this analysis by extending the waveform model up to $Q=6$ -- out of its calibration region -- obtaining similarly strong evidence of ${\cal{O}}_{C/R} = 47$. As a final check, we obtain values close to those in~\cite{Greg} when using mass-ratio priors uniform in $q\leq1$. Finally, in Table~\ref{tab:concidence} in Appendix~\ref{sec:loudness}, we report the values of individual overlap integrals for the distance and the angular sky-location (RA, DEC). \\

\paragraph{\textbf{Physical interpretation}} BBHs can efficiently merge in AGN disks as a result of gas torques and dynamical encounters \cite{2014MNRAS.441..900M,2017ApJ...835..165B,2017MNRAS.464..946S}. EM radiation must be produced in such mergers and may be detectable even against bright AGN disks, only if such disks are very thin and have relatively low luminosity or if the merger product is kicked out of the optically thick mid-plane and accretes at highly super-Eddington rates while also generating a jetted outflow \cite{2019ApJ...884L..50M,flare}. Keplerian orbital velocities for BBH in AGN disks span $O(10^{[3,4]}{\rm km/s})$, so modest kicks at an angle to BBH orbital angular momentum can lead to significant orbital perturbation and emergence from the disk, but not escape from the nucleus. From~\cite{flare}, constraints on a BBH merger come from flare start and end times, flare luminosity, and color changes (if any). 

BBH mergers at time $t_{\rm GW}$ will tend to occur in the optically thick mid-plane. For typical disk models, a flare can only be observed if the kicked merger product is in the optically thin disk atmosphere. So, the flare begins after a flight through the optically thick parts of the disk. Flare start time, therefore, depends on kick velocity, disk scale height, and the resultant of the orbital velocity ($v_{\rm orb}$) and the kick velocity ($v_{\rm kick}$), which in turn depends on the merger location in the disk. The flare ends at $t_{\rm end}$ after the mass of gas within the sphere of influence of the remnant drops below the amount required to supply the large mass accretion rate required to power the flare. Total distance travelled is $v_{\rm kick}(t_{\rm GW}-t_{\rm end})$, and assuming a mid-plane merger, the merger product leaves the merger site at angle $\theta=\sin^{-1}(v_{\rm kick}/v_{\rm orb})$

 We can also reasonably assume that a BBH formed due to migration within the disk plane will form with its orbital angular momentum parallel or anti-parallel to the local plane of the AGN disk. As the BBH semi-major axis shrinks, there is a competition between gas torques which tend to maintain the binary orbit in a co-planar orientation and dynamical interactions with the spherical nucleus component, which can drive the BBH out of the plane~\cite{2012MNRAS.425..460M}. Once the orbit is small enough that gas hardening torques become negligible, other torques (notably GR-induced) can cause precession of the binary orbit plane, so the orbit may not be precisely co-planar.


The re-interpretation of GW190521 presented here suggests several interesting possibilities for the EM flare association. First, the host AGN is unobscured and optically bright and therefore likely viewed at an angle $\lesssim 60^\circ$ to the observer (i.e., a Type I AGN). In contrast, when restricting to the sky-location of the AGN, we find that the orbit, which should be nearly aligned with the AGN, is viewed at $\iota=80^{+20}_{-36}$\,deg. Such viewing angle suggests either that the inner AGN disk itself is warped \cite{1975ApJ...195L..65B} and that the outer disk is misaligned at $\sim 80^\circ$ to line-of-sight, or that strong dynamical encounters with the spherical nuclear component are common \cite{2018MNRAS.474.5672L,2021arXiv211103992V}, most likely in a short-lived ($<5$\,Myr) dense AGN disk.
Second, the larger mass inferred here for $m_1$ in GW190521 lends greater support to an AGN origin for this merger. A primary mass $>10^{2}\,M_{\odot}$ is itself plausibly a fourth-generation BH from hierarchical mergers, and such mergers can only occur in galactic nuclei, most likely AGN~\cite{2021arXiv210903212F}.\\

\begin{figure}
\begin{center}
\includegraphics[width=0.49\textwidth]{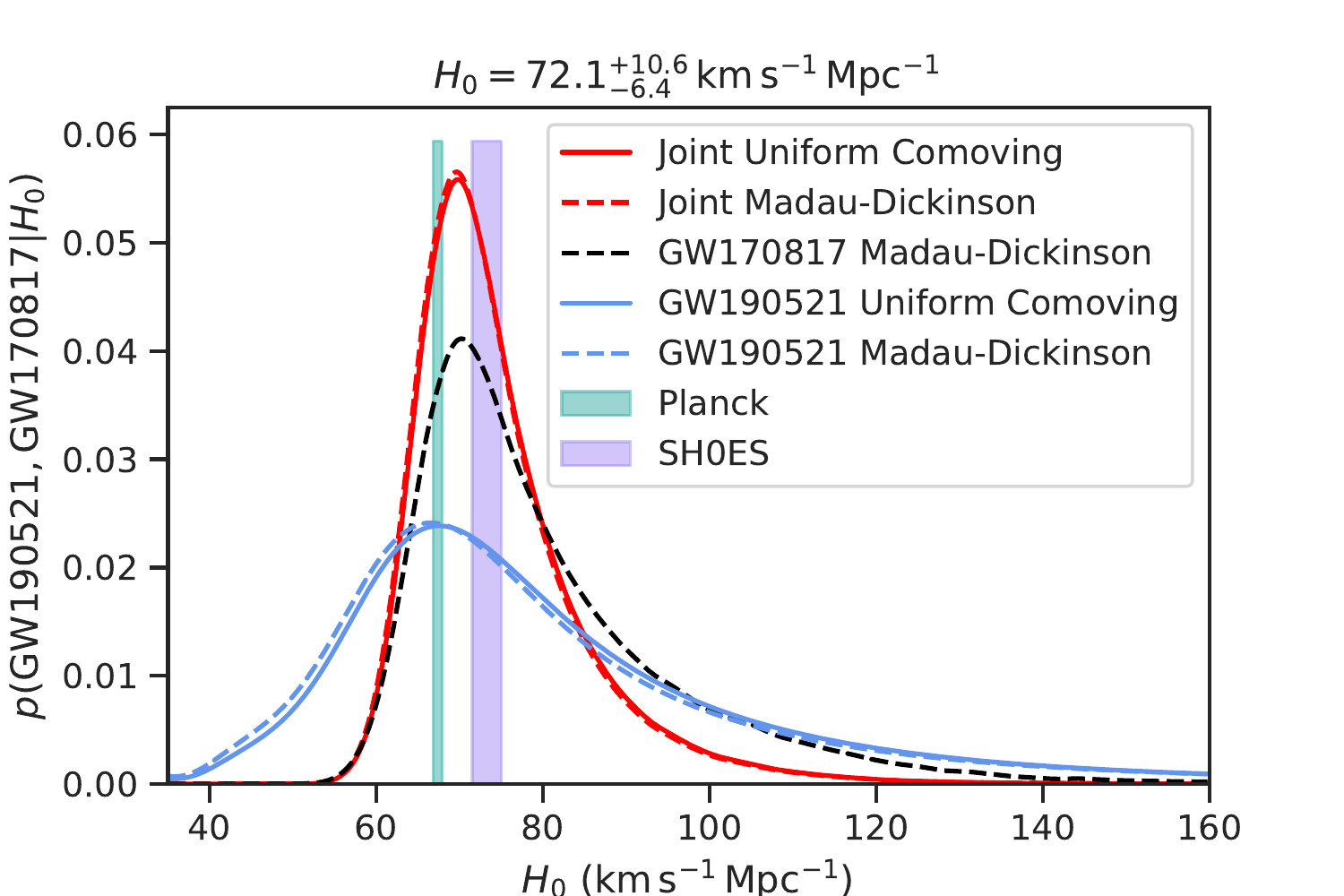}
\caption{\textbf{Hubble constant from GW170817 and GW190521}. We show results assuming a constant formation rate up to a red-shift $z=10$ and a Madau-Dickinson star-formation rate~\cite{Madau:StarFormation}. The dashed curves denote estimates obtained solely with GW190521.}
\label{fig:H0}
\end{center}
\end{figure}

\paragraph{\textbf{Hubble constant}} The expansion rate of the Universe is parametrised by the Hubble constant H$_0$, which relates the distance of a source with its recession velocity from Earth. Because GW and EM waves respectively provide distance and redshift estimates, the joint observation of a source enables an estimate of H$_0$~\cite{Schutz1986} independent of those purely based on EM information~\cite{Planck_2020,Riess_H0,Riess_2021}. This was first performed with the joint observation of the neutron-star merger GW170817-GRB170817A \cite{GW170817,Goldstein2017,Savchenko2017}, yielding a value H$_0=70.0^{+12.0}_{-8.0}\,\mathrm{km\,s^{-1}\,Mpc^{-1}}$~\cite{H0_nature}. Also e.g. \cite{H0_HsinYu,Suvodip_H0,Mastrogiovanni2021} performed estimates of H$_0$ using GW190521-ZTF19abanrhr based on the LVK parameter estimates for GW190521 \cite{GW190521I}. Following the methods in~\cite{H0_nature,H0_HsinYu}, we perform this measurement using our results for GW190521 \footnote{To compute the likelihood for the distance conditioned to the sky-location of ZTF19abanrhr $p(d_L\mid\theta_\mathrm{ZTF})$ we fit the posterior probability density $p(\mathrm{RA},\mathrm{DEC},d_L)$ with a Gaussian Mixture Model (GMM) and compute $p(\mathrm{RA}_{ZTF},\mathrm{DEC}_{ZTF},d_L)$. We re-weight by the appropriate distance prior when needed$^{4}$.}\textsuperscript{,}\footnote{For the LVK runs, the prior $\pi(d_L)\propto d_L^{2}$ is removed by the projection along the line-of-sight and no re-weighting is needed. For our runs, we re-weight by $d_L^2/\pi(d_L)$ where $\pi(d_L)$ is our prior uniform in co-moving volume}. We consider GW selection effects according to two different BBH merger rates: one constant up to a redshift $z=10$ and one following the Madau-Dickinson star-formation rate $\psi(z) = \frac{d^2 N}{dV dt} \propto \frac{(1+z)^{2.7}}{1+(1+z/2.9)^{5.6}}$ \cite{Madau:StarFormation}. For both cases, we obtain H$_0=73^{+24}_{-15}\,\mathrm{km\,s^{-1}\,Mpc^{-1}}$ at the $68\%$ credible level. This is in sharp contrast to the value of H$_0=61^{+17}_{-20}\,\mathrm{km\,s^{-1}\,Mpc^{-1}}$  obtained from the LVK results. Combining this with the kilonova GW170817-GRB170817A yields a joint estimate of H$_0=72.1^{+10.6}_{-6.4}\,\mathrm{km\,s^{-1}\,Mpc^{-1}}$, shown in Fig.\ref{fig:H0}. This implies a $15\%$ uncertainty improvement over the GW170817-GRB170817A estimate. \\ 

\begin{table*}
\centering
\begin{tabular}{l|cc|cc|cc}
Parameter  & NRsur7dq4 (Q) & NRsur7dq6 (Q) & NRsur7dq4 (q) & NRsur7dq6 (q) & LVK Settings & LVK  \\
\hline

\rule{0pt}{3ex}%
Primary mass $[M_\odot]$ & $ 106^{+36}_{-28} $ & $104^{+44}_{-26} $ & $85^{+22}_{-13}$ & $88^{+24}_{-14}$ & $85^{+19}_{-13}$ & $85^{+21}_{-14}$   \\
\rule{0pt}{3ex}%
Secondary mass $[M_\odot]$ & $ 61^{+21}_{-19} $  & $ 61^{+25}_{-24} $ & $ 70^{+20}_{-18} $ & $ 73^{+19}_{-19} $  & $68^{+20}_{-18}$ & $66^{+17}_{-18}$   \\
\rule{0pt}{3ex}%
Total mass $[M_\odot]$ & $ 168^{+29}_{-27} $ & $ 169^{+30}_{-27} $  & $ 156^{+38}_{-18} $ & $ 160^{+40}_{-22} $  & $ 152^{+30}_{-16} $   & $ 150^{+29}_{-17} $  \\
\rule{0pt}{3ex}%
Total redshifted mass $[M_\odot]$  & $255^{+32}_{-30}$    & $258^{+37}_{-37}$ & $272^{+24}_{-24}$ & $272^{+22}_{-24}$ & $274^{+25}_{-25}$   & $272^{+26}_{-27}$  \\
\rule{0pt}{3ex}%
Mass ratio &  $1.71^{+1.50}_{-0.65}$  & $1.66^{+2.00}_{-0.60}$  & $1.19^{+0.61}_{-0.18}$  & $1.18^{+0.54}_{-0.16}$  &  $1.22^{+0.65}_{-0.20}$ &  $1.26^{+0.73}_{-0.24}$ \\
\rule{0pt}{5ex}%
$P(m_1 > \text{PISN})$ &  0.15  & 0.21  & 0  & 0  & 0.001 & 0.003  \\
\rule{0pt}{3ex}%
$P(m_2 < \text{PISN})$ & 0.65 &  0.58  & 0.30 & 0.25 & 0.35 & 0.46  \\
\rule{0pt}{3ex}%
$P(m_1 > \text{PISN} \And m_2 < \text{PISN} )$  & $ 0.13 $ & 0.20  & 0 & 0  & 0  & 0 \\
\rule{0pt}{5ex}%
Primary spin  & $0.81^{+0.16}_{-0.52}$    & $0.85^{+0.12}_{-0.61}$  & $0.76^{+0.22}_{-0.62}$ & $0.73^{+0.24}_{-0.64}$  & $0.74^{+0.23}_{-0.62}$   & $0.67^{+0.29}_{-0.59}$   \\
\rule{0pt}{3ex}%
Effective spin ($\chi_\text{eff}$) & $-0.13^{+0.37}_{-0.33}$    & $-0.13^{+0.39}_{-0.37}$  & $0.03^{+0.28}_{-0.30}$ & $0.00^{+0.27}_{-0.30}$  & $0.06^{+0.28}_{-0.30}$   &  $0.08^{+0.27}_{-0.36}$ \\
\rule{0pt}{3ex}%
Effective precessing spin ($\chi_{\text{p}}$) &   $0.72^{+0.22}_{-0.32}$   &  $0.75^{+0.20}_{-0.34}$  & $0.72^{+0.22}_{-0.36}$ & $0.74^{+0.20}_{-0.35}$  & $0.73^{+0.21}_{-0.34}$  & $0.68^{+0.25}_{-0.37}$  \\
\rule{0pt}{5ex}%
Luminosity distance [Gpc] & $2.8^{+3.4}_{-1.7}$    & $2.9^{+3.3}_{-1.9}$ & $4.6^{+2.3}_{-2.5}$ & $4.3^{+2.5}_{-2.4}$  & $5.1^{+2.3}_{-2.3}$ & $5.3^{+2.4}_{-2.6}$ \\
\rule{0pt}{3ex}%
Redshift & $ 0.49^{+0.44}_{-0.26}$    & $ 0.50^{+0.43}_{-0.30}$ & $ 0.73^{+0.29}_{-0.34}$  & $ 0.70^{+0.31}_{-0.33}$  & $ 0.80^{+0.27}_{-0.32}$  & $ 0.82^{+0.28}_{-0.34}$ \\
\rule{0pt}{3ex}%
Inclination ($\theta_\mathrm{JN}$) [deg] & $56^{+30}_{-37}$    & $57^{+30}_{-38}$ &  $38^{+28}_{-25}$ & $40^{+25}_{-26}$  & $34^{+26}_{-22}$ & $31^{+28}_{-21}$ \\
\rule{0pt}{3ex}%
Orbital Inclination ($\iota$) [deg] & $65^{+21}_{-47}$   & $62^{+26}_{-44}$ &  $37^{+27}_{-26}$ & $39^{+26}_{-27}$  & $34^{+27}_{-24}$ & $31^{+29}_{-22}$ \\
\rule{0pt}{3ex}%
\rule{0pt}{5ex}%
Final redshifted mass $[M_\odot]$& $245^{+27}_{-25}$    & $247^{+31}_{-31}$ & $258^{+20}_{-21}$ & $258^{+28}_{-21}$   & $260^{+20}_{-21}$  & $258^{+22}_{-24}$  \\
\rule{0pt}{3ex}%
Final mass $[M_\odot]$ & $162^{+29}_{-28}$    & $162^{+29}_{-28}$ & $148^{+36}_{-18}$ & $151^{+39}_{-21}$   & $144^{+29}_{-16}$  & $142^{+28}_{-16}$ \\
\rule{0pt}{3ex}%
Final spin & $0.65^{+0.12}_{-0.23}$ & $0.66^{+0.12}_{-0.20}$  & $0.71^{+0.08}_{-0.10}$ & $0.71^{+0.08}_{-0.10}$  & $0.72^{+0.08}_{-0.10}$   & $0.72^{+0.09}_{-0.12}$  \\
\rule{0pt}{6ex}%


\rule{0pt}{5ex}%
$\text{Coincidence odds}^{\text{GW190521}}_{\text{ZTF19abanrhr}}$ & $72.0$   & $47.0$  & $3.4$ & $3.9$ & 2.9 & $2.4$   \\
\rule{0pt}{3ex}%
$H_0^{\text{ZTF19abanrhr}}$ & $73^{+24}_{-15}$   & $67^{+23}_{-13}$  & $67^{+18}_{-24}$ &  $49^{+19}_{-11}$ & $54^{+32}_{-15}$ & $61^{+17}_{-20}$  \\
$H_{0_,\text{GW170817}}^{\text{ZTF19abanrhr}}$ & $72.1^{+10.5}_{-6.4}$  & $70.7^{+9.8}_{-5.9}$  & $71.2^{+9.0}_{-6.1}$ & $69.8^{+8.1}_{-5.8}$ &  $72.4^{+12.3}_{-7.2}$ & $70.2^{+8.2}_{-5.7}$  \\
\rule{0pt}{5ex}%
Maximum SNR & 15.90    & 15.83  & 15.72 & 15.77  & 15.57 & 15.42    \\
\rule{0pt}{3ex}%
Maximum LogL ($\ln{\mathcal{L}_\textrm{max}}$) &  119.2  &  117.5 & 116.5 & 117.1 & 114.1 & --   \\
\rule{0pt}{3ex}%
LogBayes Factor ($\ln\mathrm{BF}^{\rm S}_{\rm N}$)&  88.79    & 88.43  & 88.65 & 88.85 & 88.66  & --  \\

\end{tabular}
\caption{\textbf{Summary of GW190521 parameters for our different runs.} We quote median values and symmetric $90\%$ credible intervals except for H$_0$ values, for which we quote $68\%$ ones. The first two columns make use of a mass-ratio prior uniform in $Q\in[1,4]$ and $Q\in[1,6]$ while the next two use a prior uniform in $q\in[1/4,1]$ and $q\in[1/6,1]$.
We use a flat prior on the total red-shifted mass. The \texttt{NRSur7dq4} model is calibrated up to mass ratios of 4 but can be extrapolated up to mass ratios of 6. Rows 5-7 assume a PISN gap spanning the mass range $[65-130]M_\odot$. The inclination angles $\theta_{JN}$ and $\iota$ between the line of sight and, respectively, the total angular momentum and the orbital angular momentum are quoted in terms of $\alpha_{\text{quoted}}=\pi/2 - |\pi/2-\alpha|$ so that $\pi/2$ corresponds to edge-on orientation and $0$ corresponds to both face-on and face-off. These, together with the spin values are quoted at a reference frequency $f_\text{ref}=11$\,Hz. The last two columns show the results obtained on the posterior samples released by the LVK and our reproduction of the LVK run using their prior settings, namely a mass-prior uniform in red-shifted component masses and a distance prior $p(d_L)\propto d_L^2$. The original LVK analysis was performed with the LALInference \cite{Veitch:2015ela} algorithm using a nested sampler and 2048 live points, while ours were performed with \texttt{Bilby} using the \texttt{Dynesty}~\cite{Dynesty} sampler and a more aggressive 4096 live points.}
\label{tab:PE}
\end{table*}

{\it \textbf{Discussion.}} 
GW190521 has challenged the fields of gravitational-wave astronomy and astrophysics. Its ``simplest'' interpretation of being a quasi-circular BBH required waveform models not available at the time of its detection \cite{NRSur7dq4,Babak:2016tgq,SEOBNRv4PHM,PhenomPv3HM,XPHM_Pratten} and poses puzzles such as a merging BH populating the PISN gap. Also, its short signal length makes its interpretation very sensitive to prior assumptions. Along the lines of \cite{GW190521_Nitz,Hector_GW190521} we have shown that usage of standard prior assumptions in Bayesian inference can highly suppress parameter regions of high likelihood.

Removing such suppression, we have shown that GW190521 is consistent with a quasi-circular BBH with a larger mass ratio, larger inclination, and much closer distance than originally reported by the LVK. More importantly, the sky location of GW190521 is consistent with that of its proposed electromagnetic counterpart ZTF19abanrhr. We obtain an odds of ${\cal{O}}_{C/R}=71$ favouring a true coincidence in high contrast the ${\cal{O}}_{C/R}=2.8$ reported for the original LVK results \cite{Greg,GW190521D}. This result points, for the first time, to GW190521 as a strong candidate for the first multi-messenger observation of a compact merger with masses in the black-hole range.

An association between GW190521 and an EM counterpart leads to several astrophysical constraints. First, in order to explain the flare, the kicked, merged BH must be accreting at super-Eddington rates, with radiation escaping in a jet. Second, the orientation of the binary implies that \emph{either} the outer disk of the AGN (where the merger occurred) is strongly warped compared to the inner disk, \emph{or} dynamical interactions with the spheroidal component are common. The latter is more likely in a short-lived, dense AGN disk. Third, the higher primary mass combined with a break in the LVK observed mass function around $35\,M_{\odot}$ is suggestive of a high generation primary that could only merge in a galactic nucleus.   

Finally, we note that this study is limited to the ``standard'' scenario of a quasi-circular BBH. However, GW190521 has also been shown to be consistent an eccentric merger \cite{Isobel_ecc,Gayahtri_ecc,GW190521_Gamba} of BBHs and even an exotic head-on merger of Proca-stars \cite{Proca}. The future development of waveform models accounting for both precession and eccentricity shall yield further clues on the true nature of GW190521 and the robustness of our results.
 
\section*{Acknowledgements}
We thank Will Farr, and Greg Ashton for the respective public release of their codes to compute the value of the Hubble constant \cite{Will_notebook} and the coincidence odds \cite{Greg} as well as for valuable discussions. We also thank Hector Estelles, Sergei Ossokine, and Vijay Varma for their help to understand the impact of possible waveform systematics and Leo Singer for his advice on the usage of his ClusteredKDE code \cite{Leo_KDE} to compute the values in Table~\ref{tab:concidence}.
The analyzed data is publicly available at the online Gravitational Wave Open Science Center \cite{GWOSC}.
LVK results quoted throughout the paper and the corresponding histograms and contours have made use of the publicly available sample release in \texttt{https://dcc.ligo.org/P2000158-v4}.
JCB is supported by a fellowship from ``la Caixa'' Foundation (ID 100010434) and by the European Union’s Horizon 2020 research and innovation programme under the Marie Skłodowska-Curie grant agreement No 847648. The fellowship code is LCF/BQ/PI20/11760016. JCB is also supported by the research grant PID2020-118635GB-I00 from the Spain-Ministerio de Ciencia e Innovaci\'{o}n. This work has received financial support from Xunta de Galicia (Centro singular de investigaci\'{o}n de Galicia accreditation 2019-2022), by European Union ERDF, and by the “Mar\'{i}a de Maeztu” Units of Excellence program MDM-2016-0692 and the Spanish Research State Agency. KC acknowledges the MHRD, Government of India, for the fellowship support. SHWL is partially funded by the Department of Physics at The Chinese University of Hong Kong. BM and KESF are supported by NSF AST-1831415 and Simons Foundation Grant 533845. We acknowledge the use of IUCAA LDG cluster Sarathi for the
computational/numerical work. The authors acknowledge computational resources provided by the CIT cluster of the LIGO Laboratory and supported by National Science Foundation Grants PHY-0757058 and PHY0823459, and the support of the NSF CIT cluster for the provision of computational resources for our parameter inference runs. This material is based upon work supported by NSF's LIGO Laboratory which is a major facility fully funded by the National Science Foundation. This manuscript has LIGO DCC number P2100467. 

\appendix

\begin{figure*}
\centering
\includegraphics[width=0.32\textwidth]{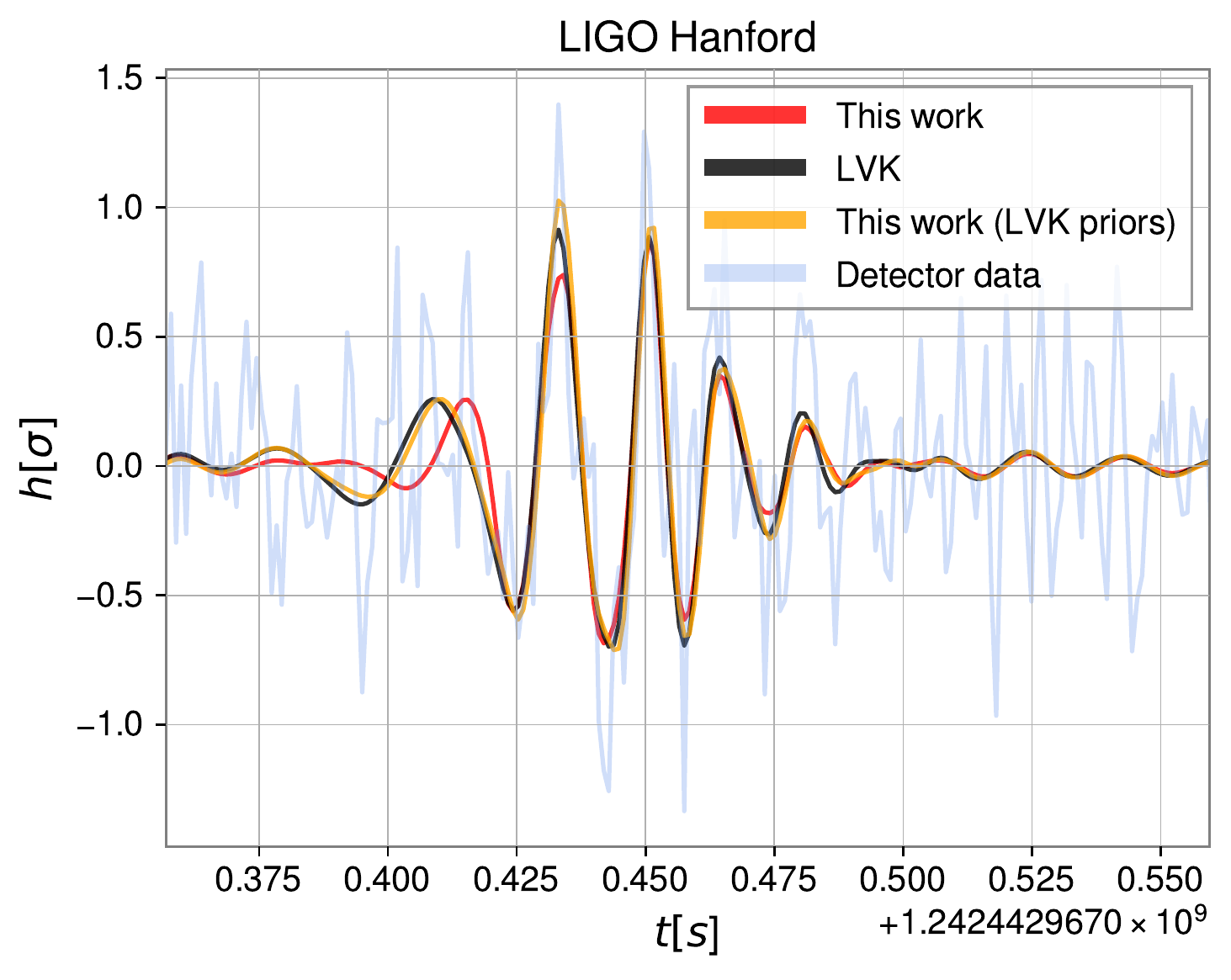}
\includegraphics[width=0.32\textwidth]{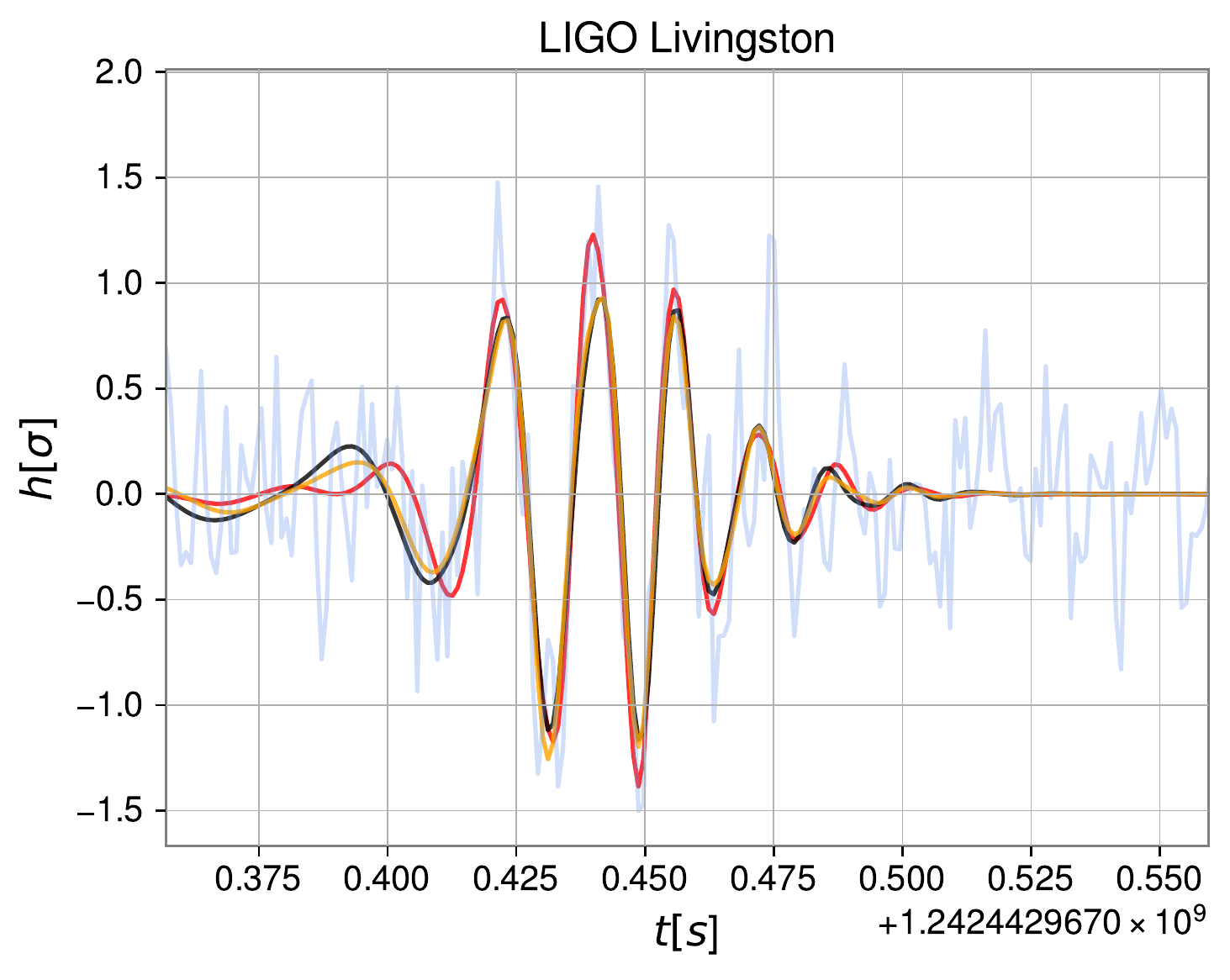}
\includegraphics[width=0.32\textwidth]{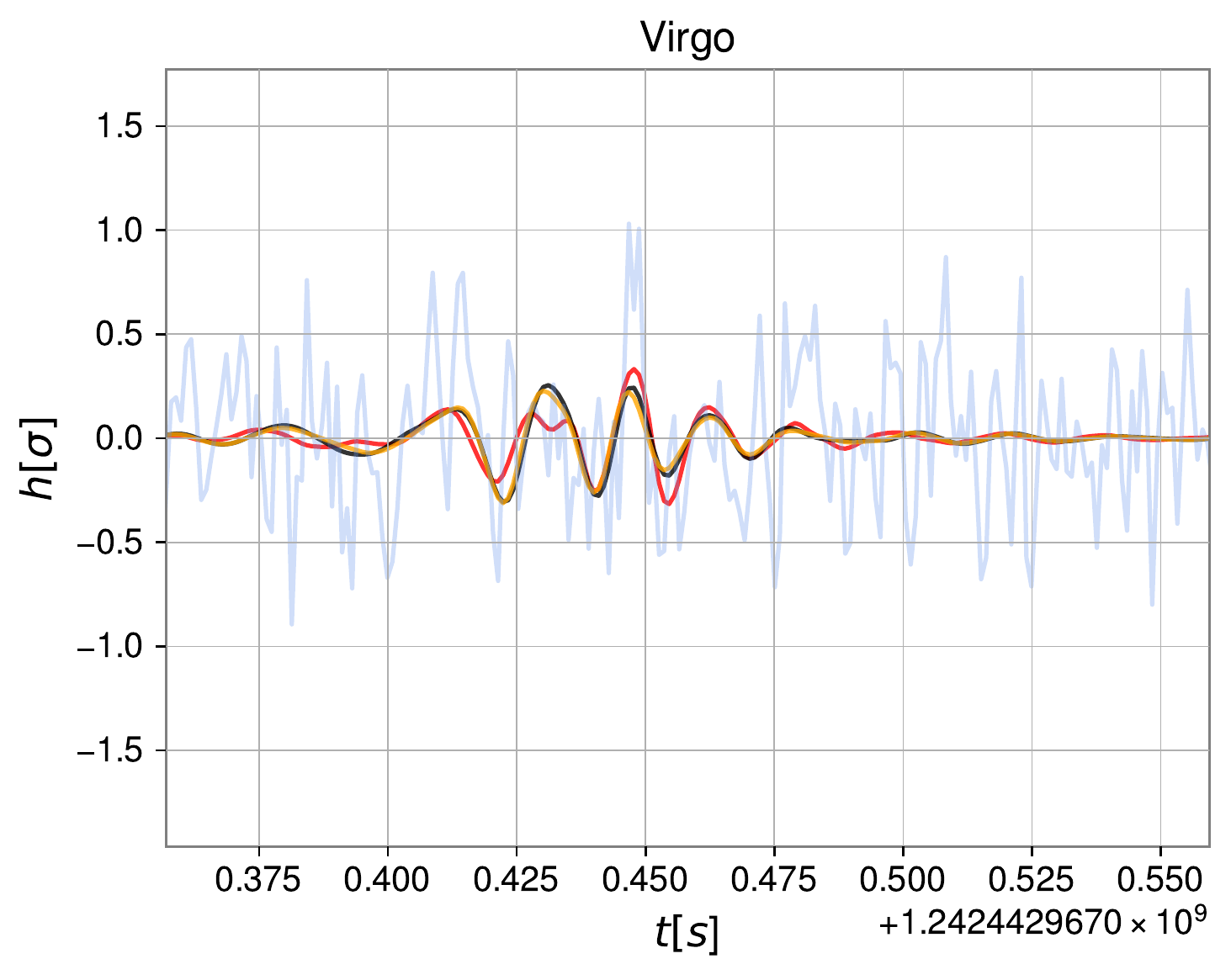}
\caption{\textbf{Maximum likelihood waveforms for GW190521 according to different analyses}. Whitened maximum likelihood (best-fitting) waveforms for GW190521 obtained by this work (red), the LVK (black) and by our reproduction of the LVK analysis (orange), overlaid with the corresponding detector strain data.}
\label{fig:waveforms}
\end{figure*}

\section{Maximum Likelihood waveforms}
\label{sec:maxL}

Fig.\ref{fig:waveforms} shows the whitened maximum likelihood waveforms corresponding to the three analyses shown in Fig.\ref{fig:masses} in the main text together with the corresponding detector strain data. We show our maximum likelihood waveform in red while those in black and orange correspond to those of the LVK analysis and our reproduction of it.

We note two main aspects. First, while the black and yellow waveforms overlay well during the whole signal duration, ours differ visibly in the Hanford and Livingston cases in two main aspects. These are a steeper signal rise before the merger and a different height at the signal peaks. Second, for the case of Virgo, our waveform shows some complex structure (see between $t=0.425$ and $t=0.450$) that is not present in the other two. These differences, visible by eye, are behind the highly differing maximum likelihood values reported in Table~\ref{tab:PE}. 



\subsection{Comparison with Nitz et. al and Estelles et. al.}

\subsubsection{Phenomenological models}

As noted in the main text, \citet{GW190521_Nitz} and \citet{Hector_GW190521} did also analyse GW190521 using a mass prior uniform in mass ratio $Q\geq 1$ but reached a different conclusion on the potential coincidence with ZTF19abanrhr. The analysis of \cite{Hector_GW190521} and part of that of \cite{GW190521_Nitz} where conducted with  phenomenological waveform models that are not calibrated to numerical simulations of precessing black holes and therefore may not accurately account for all the physics in such systems. Instead, these waveform models are calibrated to simulations of non-precessing binaries and precession is then accounted for by ``twisting'' the resulting waveforms using analytical expressions whose validity breaks during the merger and ringdown regimes. In addition, for instance, these models assume a conjugation symmetry $h_{\ell,-m} = (-1)^{m}h_{\ell,m}^{*}$ between right-handed and left-handed GW emission modes which is not present in precessing systems. For these reasons, we argue that these models should not be taken as a reference when analysing a system like GW190521 for which precession is a relevant effect and for which higher-harmonics can play an important role due to its high mass \cite{Graff:2015bba,Capano:2013raa,Bustillo:2016gid,Pang:2018hjb,Harry:2017weg,Chandra2020}.\\

\begin{figure}
\centering
\includegraphics[width=0.49\textwidth]{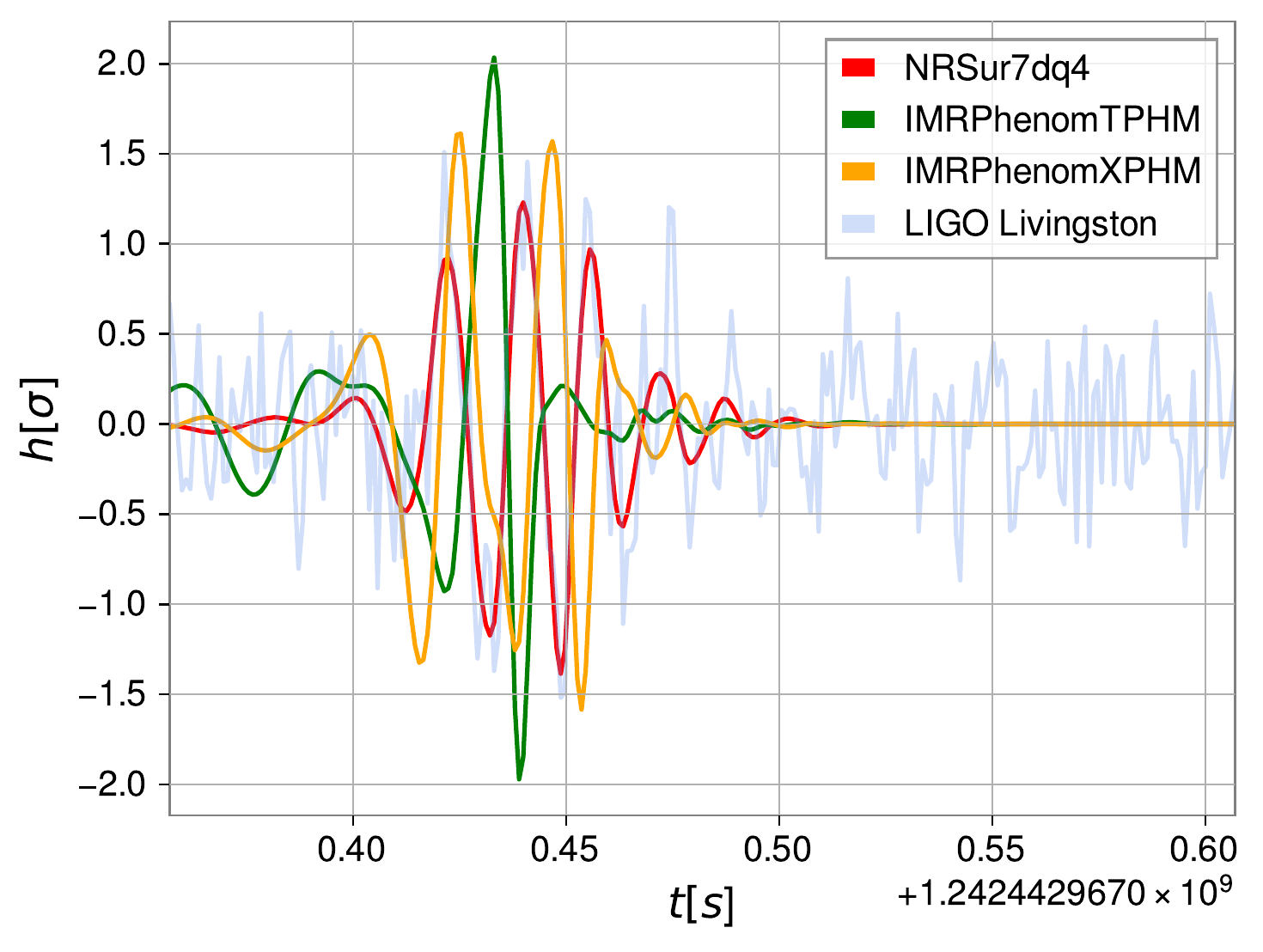}
\caption{\textbf{Comparison of the different waveform models at the best-fit parameters reported by this work.} Whitened waveforms computed by the waveform model employed in this work (\texttt{NRSur7dq4}) and the models  \texttt{IMRPhenomTPHM} and \texttt{IMRPhenomXPHM} respectively employed \citet{GW190521_Nitz} and \citet{Hector_GW190521}, for the maximum likelihood parameters we obtain from \texttt{NRSur7dq4} when using a mass-ratio prior uniform in $Q\in[1,4]$.}
\label{fig:phenom}
\end{figure}

As a cross-check, we performed our reproduction of the above two studies by repeating our runs using the models \texttt{IMRPhenomXPHM} used by \citet{GW190521_Nitz} and \texttt{IMRPhenomTPHM} used by \citet{Hector_GW190521}. We did this for mass ratio priors uniform in $Q\in[1,4]$ and $Q\in[1,6]$ (to allow for an apples-to-apples comparison with our results) and for $Q\in[1,10]$. In all cases, we obtain weak evidence for a true coincidence with ZTF19abanrhr. Finally, we showcase the vastly different predictions that these models make at the points in the parameter space we are exploring. Fig.~\ref{fig:phenom} shows the waveforms computed with \texttt{NRSur7dq4}, \texttt{IMRPhenomTPHM} and \texttt{IMRPhenomXPHM} at the maximum likelihood parameters values reported by \texttt{NRSur7dq4} together with the corresponding Livingston strain data. It is clear that both phenomenological models predict extremely different waveforms. Consequently, it is not surprising that they lead to very different conclusions.

\subsubsection{\texttt{NRSur7dq4}}

As in this work, \citet{GW190521_Nitz} did also compare GW190521 to the model \texttt{NRSur7dq4}. In doing so, however, they reached mass ratios of $Q=6$ beyond the calibration region $Q\leq 4$ of the model, finding a region of high probability density at mass-ratios $Q\geq 4$. As they argued, resolving such features requires the usage of live-points that safeguard the analysis against mode “die-off”, which may prevent the parameter sampler from ignoring certain regions of the parameter space if these are not populated by live points early enough in the analysis. While typical analyses, including the one we present here, use less than 5k live points, their analysis used up to 40k. As a modest cross-check, we repeated our $Q\leq 6$ run with 8096 live points, finding two notable features.\\

\begin{figure}
\centering
\includegraphics[width=0.49\textwidth]{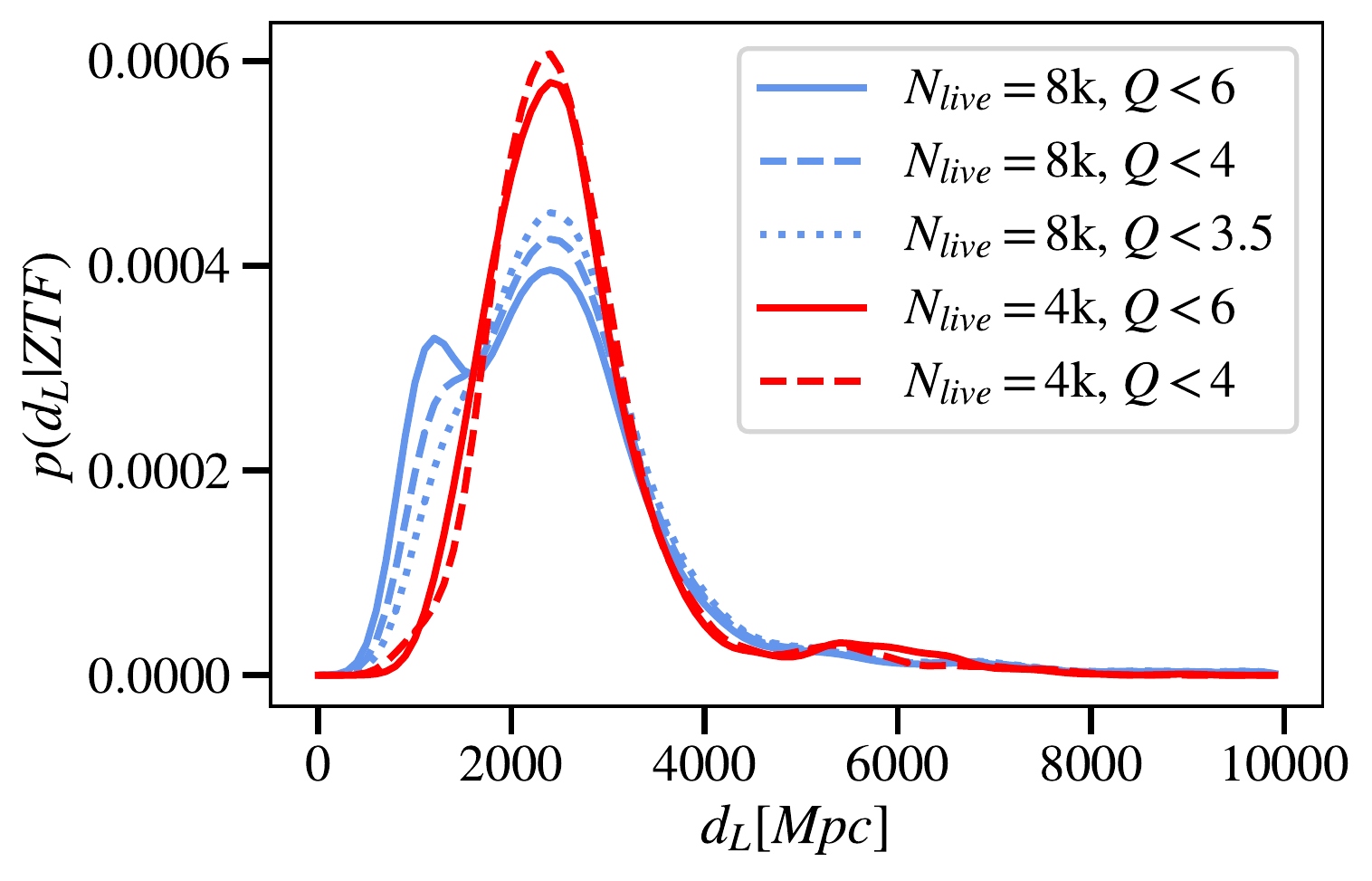}
\caption{\textbf{Impact of live points on distance estimates conditioned to the location of ZTF19abanrhr}. We show posterior distributions for the distance of GW190521 conditioned to the sky-location of ZTF19abanrhr for analyses conducted with different number of live points and for different cuts in the mass-ratio Q. }
\label{fig:8k}
\end{figure}

First, while we do not recover a peak in the mass-ratio distribution at $Q>4$ (most likely due to our still lower number of live points) but we find that the maximum likelihood point lies beyond $Q = 4$. Second, when restricting the sky-location to ZTF19abanrhr, we find a bi-modal distribution with a strong peak at $d_L \simeq 1000$ Mpc that is not present in 4k-live-points analysis presented in the main text. We show this feature in Fig.\ref{fig:8k} in the blue-solid curve. The dashed and dotted curves show that this feature mostly comes from the beyond-calibration $Q>4$ region. We note, nevertheless, that even in the presence of this secondary peak, we obtain an odds ratio of $50:1$ in favour of the coincidence hypothesis, consistent with the $47:1$ we obtain using 4k live points.

\section{\label{sec:overlap}Detailed coincidence overlap integrals}

In the main text we have reported the full three-dimensional odds-ratio ${\cal{O}}_{C/R}$ for a true coincidence between GW190521 and ZTF19abanrhr, computed as ${\cal{O}}_{C/R}={\cal{I}}_\theta \times \pi_{C/R}$ \cite{Greg}. Here, $\pi_{C/R}$ denotes the prior odds while ${\cal{I}_{\theta}}$ is known as the overlap integral. This is equal to the ratio between the probability for the distance and sky-location $\theta = (d_L^{\text{ZTF}},\Omega^{\text{ZTF}})$ of ZTF19abanrhr given the corresponding posterior distributions $p(d_L,\Omega)$ obtained from the analysis of the GW190521 data $d_{\text{GW190521}}$; and the corresponding prior probability $\pi(d_L,\Omega)$. In particular one can express:

\begin{equation}
    \mathcal{I}_\theta = \int \frac{p(\theta \mid d_\text{GW190521},\,C)\,p(\theta \mid d_\text{ZTF},\,C)}{\pi(\theta \mid C)}\,d\theta.
\end{equation}

Since the distributions for the ZTF parameters can be considered as delta functions, we can simplify the above to:

\begin{equation}
    {\cal{I}}_\theta \equiv {\cal{I}}_{d_L,\Omega} = \frac{p(d_L^{\text{ZTF}},\Omega^{\text{ZTF}}\mid d_{\text{GW190521}})}{\pi(d_L^{\text{ZTF}},\Omega^{\text{ZTF}} \mid C)},\\
\end{equation}
    and assuming $d_L$ and $\Omega$ are independent, 
\begin{align}
    {\cal{I}}_{d_L,\Omega} &\simeq \frac{p(d_L^{\text{ZTF}}\mid d_{\text{GW190521}})}{\pi(d_L^{\text{ZTF}}\mid C)} \frac{p(\Omega^{\text{ZTF}} \mid d_{\text{GW190521}})}{\pi(\Omega^{\text{ZTF}}\mid C)} \\
    &=  {\cal{I}}_{d_L} {\cal{I}}_{\Omega} \nonumber.
\end{align}

Here $({\cal{I}}_{d_L},{\cal{I}}_{\Omega})$ denote the ``partial'' overlap integrals. For a detailed derivation, please see \cite{Greg}. In Table~\ref{tab:concidence} we show the values of these integrals, together with the full three-dimensional ones of the various analyses presented in this work. It is evident that the $Q$-analyses report values for the individual integrals are larger than those of the remaining analyses by factors of, respectively $\simeq 2$ and $\simeq 4$. Finally, for completeness, we re-calculated the LVK results are reported in~\cite{Greg} (see next).

\subsection{Accuracy of overlap integral ${\cal{I}}_{d_L,\Omega}$ and corresponding odd-ratios}

\begin{figure}
\centering
\includegraphics[width=0.49\textwidth]{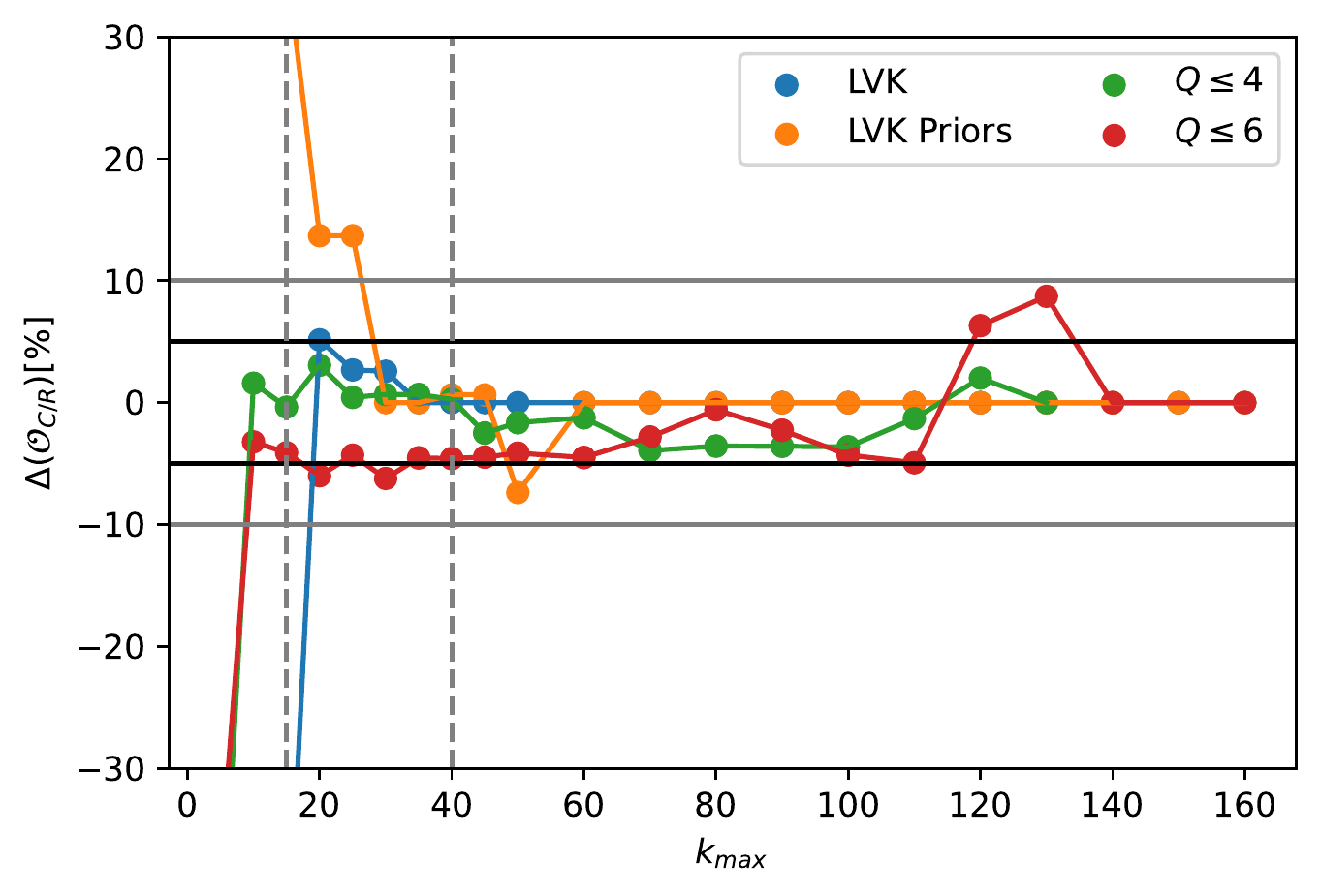}
\caption{\textbf{Dependence of coincidence odd-ratios on the accuracy of three-dimensional Kernel Density Estimates}. We show the percent difference between ${\cal{O}}_{C/R}$ computed with a given resolution of the three dimensional KDE for $p(d_L,\Omega)$ and that computed at a maximum resolution. The resolution is parametrised by the number of allowed clusters $k_{\text{max}}$ allowed to compute the KDE. The horizontal lines denote differences of $\pm 5\%$ and $\pm 10\%$.}
\label{fig:kde}
\end{figure}

The calculation of ${\cal{I}}_{d_L,\Omega}$ involves an estimation of the joint three-dimensional probability density distribution for the sky-angles and luminosity distance $p(d_L,\Omega)$. To this, both the authors of \cite{Greg,Hector_GW190521} and us used the multi-dimensional Kernel Density Estimator (KDE) \texttt{ClusteredKDE} described in \cite{Leo_KDE}. A crucial element of such KDE estimation is the maximum number of clusters $k_{\text{max}}$ that can be defined within the sample set. The algorithm finds then an optimal number of clusters $k<k_{\text{max}}$ that maximises the Bayesian Information Criterion, or BIC. To ensure the robustness of our results, we computed our ${\cal{I}}_{d_L,\Omega}$ (or equivalently, ${\cal{O}}_{C/R}$) for varying values of $k_{\text{max}}\in[5,160]$. Fig.~\ref{fig:kde} shows the percent difference between ${\cal{I}}_{d_L,\Omega}$ computed for different $k_{\text{max}}$ and $k_{\text{max}}=160$. The two pairs horizontal of lines denote respectively denote $\pm 5\%$ and $\pm 10\%$ differences. For reference, the vertical lines denote the values $k_{\text{max}} = 15$ used in \cite{Greg} and $k_{\text{max}} = 15$ used by default in \cite{Leo_KDE}.

\begin{table}
\centering
\begin{tabular}{lcccc}
Analysis & ${\cal{I}}_{d_L}$ & ${\cal{I}}_{\Omega}$ & ${\cal{I}}_{d_L,\Omega}$  & ${\cal{O}}_{C/R}^{\text{3D}}$ \\
\hline
\rule{0pt}{3ex}%
LVK$^{\text{\citet{Greg}}}$ & 1.75   & 29.26 & 31.3  & 2.4  \\
\rule{0pt}{3ex}%
LVK & 1.75  & 29.26 &  61.1  & 4.6  \\
\rule{0pt}{3ex}%
LVK Priors          & 1.71  & 18.90  & 38.1  & 2.9  \\
\rule{0pt}{3ex}%
$Q \in [1,4]$       & 4.07  & 118.83 & 937.1 & 72.0  \\
\rule{0pt}{3ex}%
$Q \in [1,6]$        & 3.54  & 86.59  & 611.7 & 47.0   \\
\rule{0pt}{3ex}%
$q \in [1/4,1]$        & 1.63  & 12.96  & 44.2  & 3.4    \\
\rule{0pt}{3ex}%
$q \in [1/6,1]$ & 1.98  & 27.06  & 50.0  & 3.9  \\
\end{tabular}
\caption{\textbf{Detailed GW190521 vs ZTF coincidence results.} Overlap integrals for the sky-location of GW190521 its potential counterpart and odds-ratio in favour of the coincidence hypothesis. The last five rows report values for our different analyses of GW190521, that of our reproduction of it. The first two show the values reported by \cite{Greg} using the original LVK results, and the second shows our recalculation of these values.}
\label{tab:concidence}
\end{table}

We highlight two main aspects. First, we note that large variations happen for $k_{\rm max} \leq 30$. For instance, the value obtained from the LVK samples for $k_{\rm max} \leq 20$ is twice that reported in \cite{Greg}, which we label by LVK$^{\text{Ashton et. al.}}$ in Table \ref{tab:concidence}. This, nevertheless, does not change the conclusion that evidence for a true coincidence between GW190521 and ZTF19abanrhr under the LVK results is weak. Second, we show that our results vary by less than $\simeq 5\%$ for $k_{\rm max} \geq 80$ up to our reference value of $k_{\rm max} \geq 160$. The exception to this $Q\leq 6$ analysis for which there are excursions towards $+10\%$ levels that, in any case, would not change our qualitative conclusions.

\section{\label{sec:loudness}Intrinsic source loudness}
\begin{figure}[ht!]
\centering
\includegraphics[width=0.49\textwidth]{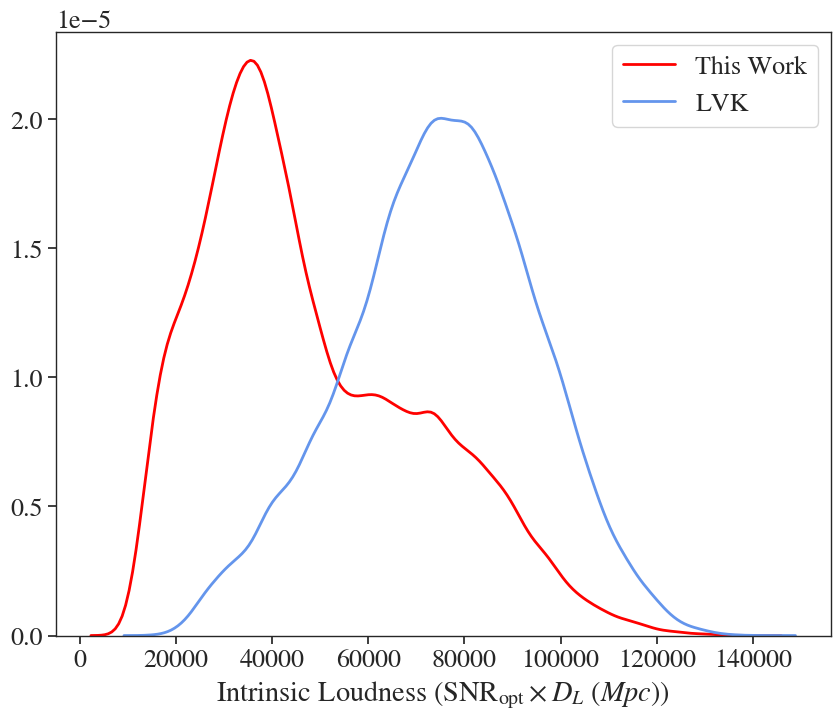}
\caption{\textbf{Posterior distribution for the intrinsic loudness of the source of GW190521.} This is computed as the product of the optimal network signal-to-noise ratio and the luminosity distance of the posterior samples. The LVK analysis strongly prefers sources that are louder by a factor of $\simeq 3$, explaining the vast difference between the corresponding distance estimates.}
\label{fig:loudness}
\end{figure}

As mentioned in the main text, the decrease in luminosity distance reported by our analysis is due to a decrease in the estimated intrinsic loudness of the sources. This is due to the fact that the LVK analysis uses a prior that favors intrinsically louder $q \approx 1$ sources. Fig \ref{fig:loudness} shows the posterior distribution of the intrinsic loudness from each analyses, computed as the product of the optimal SNR of the corresponding waveform multiplied by the corresponding estimated luminosity distance, as in \cite{[][{ (see Footnote 8) }] Chatziioannou:2019dsz}. It is clear that, in fact, our analysis shows a strong preference for weaker sources.

\clearpage
\bibliographystyle{apsrev4-1}
\bibliography{ZTF_Flare.bib}
\end{document}